\documentclass[11pt]{article}
\usepackage{amssymb}
\def\one{1\hskip -.37em 1}     
\def\proj{\mathbb E}
\def\R{\mathbb R}
\def\Z{\mathbb Z}
\begin{document}
\begin{titlepage}
\begin{centering}
 
{\ }\vspace{2cm}
 
{\Large\bf The Schwinger Model and the Physical Projector:}

\vspace{5pt}

{\Large\bf a Nonperturbative Quantization without Gauge Fixing}

\vspace{2cm}

Gabriel Y.H. Avossevou\\
\vspace{0.5cm}
{\em Unit\'e de Recherche en Physique Th\'eoriqe (URPT)}\\
{\em Institut de Math\'ematiques et de Sciences Physiques (IMSP)}\\
{\em B.P. 2628 Porto-Novo, Republic of Benin}\\
{\tt avossevou@yahoo.fr}

\vspace{0.5cm}

and

\vspace{0.5cm}

Jan Govaerts\\
\vspace{1.0cm}
{\em Institute of Nuclear Physics, Catholic University of Louvain}\\
{\em 2, Chemin du Cyclotron, B-1348 Louvain-la-Neuve, Belgium}\\
{\tt jan.govaerts@fynu.ucl.ac.be}

\vspace{1.5cm}

\begin{abstract}

\noindent
Based on the physical projector approach, a nonpertubative quantization of 
the massless Schwinger model is considered which does not require any 
gauge fixing. The spectrum of physical states,
readily identified following a diagonalization of the operator
algebra, is that of a massive pseudoscalar field, namely 
the electric field having acquired a mass proportional to the gauge
coupling constant. The physical spectrum need not be identified with confined 
bound fermion-antifermion pairs, an interpretation which one is otherwise 
led to given whatever gauge fixing procedure but 
which is not void of gauge fixing artefacts.

\end{abstract}

\vspace{25pt}
 
To be published in the Proceedings of the\\
Second International Workshop on Contemporary Problems in Mathematical
Physics,\\
Institut de Math\'ematiques et de Sciences Physiques (IMSP), Universit\'e
d'Abomey-Calavi,\\
Cotonou, Republic of Benin\\
October 28$^{\rm st}$ - November 2$^{\rm th}$, 2001

\end{centering} 

\vspace{100pt}


\end{titlepage}

\section{Introduction}
\label{Sect1}

Ever since its inception,\cite{Schwinger} the massless Schwinger model, namely
1+1 dimensional massless quantum electrodynamics, has been a favourite
theoretical laboratory of mathematical physics in which to test our
understanding and methods of interacting relativistic quantum field theories. 
This model is one of the very few instances of an integrable quantum 
field theory possessing an exact nonperturbative 
solution.\cite{Schwinger,SchM1,Manton,Hetrick} It also displays most of
the nonperturbative features of the theory for the strong interactions
of quarks, namely quantum chromodynamics (QCD) based on the nonabelian
SU(3)$_C$ colour gauge symmetry, leading to the still open
issues of confinement and spontaneous dynamical chiral symmetry breaking 
with its cort\`ege of cha\-rac\-te\-ris\-tic consequences for hadronic 
interactions.  Hence, it is not suprising that whenever a new nonperturbative 
phenomenon is suggested to occur in quantum field theory, or a new methodology
towards such issues is designed, the Schwinger model is brought
into the arena to test out new ideas. The present contribution is no exception.
Given the recent proposal\cite{Klauder} of the physical projector approach 
to the quantization of gauge invariant systems (and more generally, 
constrained ones),\cite{Gov1}
it thus seems timely to investigate the potential new insight which that
approach may bring to the nonperturbative issues characteristic of the
Schwinger model. The efficacy of the physical projector has already been
demonstrated in the case of gauge invariant systems in 0+1 
dimensions,\cite{Gov2} as well as topological quantum field theories in 2+1 
dimensions,\cite{Gov3} with their own specific circumstances. It would thus 
be welcome to demonstrate that the physical projector approach is also of 
relevance with regards to the nonperturbative quantization of interacting 
field theories.

Within a continuum spacetime, all previous 
approaches\cite{Schwinger,SchM1,Manton,Hetrick} to the Schwinger model
(SchM), in one way or another, are based in an essential manner on a choice
of gauge fixing of the U(1) local symmetry. In effect, this leads to the
reduction of the gauge boson degrees of freedom expressed in terms of the
fermionic ones, except for some gauge invariant zero mode in the case of
a compact space topology, namely a circle. The latter choice is 
made\cite{Manton,Hetrick} in order 
to render manifest the topological features of the SchM. The conclusion is
then that the spectrum of gauge invariant states is that of a massive
scalar field, whose quanta are the confined states of
fermion-antifermion pairs bound to one another by an
electric flux line. Indeed, the one-dimensional Gauss law in vacuum 
implies a constant electric field, hence a linearly rising
electric potential leading to a confining interaction. However, such an 
intuitive picture raises some difficulties of interpretation. For instance,
given two point particles of opposite charges and momenta forming a scalar
bound state on the circle, on which ``side" of the circle 
should the electric field binding these two charges be set-up?
As a matter of fact, other such artefacts arise as a consequence of
gauge fixing, which tend to confuse the physical understanding and
interpretation of the nonperturbative dynamics of the model.

Within the physical projector approach which does not require any gauge fixing,
such issues are avoided altogether.\cite{Gov1,Gov4} Physical states are 
nothing but the quanta of the electric field,\cite{Gaby} which is a 
pseudoscalar field in 1+1 dimensions, indeed the only spacetime local construct
based on the original degrees of freedom which is genuinely gauge invariant.
The fermionic degrees of freedom
contribute in a gauge invariant manner to the physical spectrum only nonlocally 
through their bosonized representation and only in combination with the 
local gauge dependent degrees of freedom of the gauge boson sector. Even though
the same physical spectrum is of course recovered as in any of the gauge fixed
quantizations, we find such a physical picture much more appealing than the 
usual one, the more so since within the Hamiltonian formulation of the SchM, 
the electric field readily finds its rightful place and is obviously a gauge 
invariant field whose quanta must belong to the physical spectrum whatever 
the status of the other degrees of freedom.

A full account of the analysis is not presented,\cite{Gaby} restricting the
discussion to its most salient features. Sect.\ref{Sect2} introduces the
model and our notations. Sect.\ref{Sect3} describes its system of constraints
and basic Hamiltonian formulation. A nonperturbative 
quantization is developed in Sect.\ref{Sect4}, based on the bosonization 
of the fermion degrees of freedom and the operator 
dia\-go\-na\-li\-za\-tion of the model. In Sect.\ref{Sect5}, the physical 
spectrum is identified through the physical projector. Finally, some 
conclusions are presented in Sect.\ref{Sect6}.

\section{The Massless Schwinger Model}
\label{Sect2}

The topology of the 1+1 dimensional spacetime is taken to be $\R\times S$,
$\R$ standing for the time coordinate, $t$, and $S$ for the spatial one, $x$, 
with the geometry of a circle of radius $R$ and circumference $L=2\pi R$. The
Minkowski metric is $\eta_{\mu\nu}=\,{\rm diag}\,(+-)$
$(\mu,\nu=1,2)$, while units such that $\hbar=1=c$ are used throughout.
The antisymmetric tensor $\epsilon^{\mu\nu}$ is such that $\epsilon^{01}=+1$.
For the Clifford-Dirac algebra $\{\gamma^\mu,\gamma^\nu\}=2\eta^{\mu\nu}$,
we shall work with the chiral representation given by
\begin{equation}
\gamma^0=\sigma^1\ \ ,\ \ \gamma^1=i\sigma^2\ \ ,\ \ 
\gamma_5=\gamma^0\gamma^1=-\sigma^3\ ,
\end{equation}
$\sigma^i$ $(i=1,2,3)$ being the usual Pauli matrices.

The field degrees of freedom of the model are, on the one hand, a single
massless Dirac spinor $\psi(x^\mu)$, and on the other hand, the real U(1) 
gauge vector field $A_\mu(x^\mu)$. The Dirac spinor decomposes into two 
complex Weyl spinors of opposite chiralities, $\psi=\psi_++\psi_-$, such that 
$\gamma_5\psi_\pm=\mp\psi_\pm$. The index ``$\pm$" refers to the left- or 
right-moving character, respectively, of these two Weyl spinors
in the absence of any interaction. Furthermore, the following choice of 
periodic and twisted boundary conditions on the circle is assumed,
\begin{equation}
A_\mu(t,x+L)=A_\mu(t,x)\ \ \ ,\ \ \ 
\psi_\pm(t,x+L)=-e^{2i\pi\alpha_\pm}\,\psi_\pm(t,x)\ ,
\end{equation}
$\alpha_\pm$ being arbitrary real quantities defined modulo any integers,
which, in fact, have to be equal modulo integers to ensure parity invariance, 
$\alpha_+=\alpha=\alpha_-$ (mod $\Z$).

The dynamics of the model derives from the local Lagrangian density
(the summation convention is implicit throughout)
\begin{equation}
{\cal L}=-\frac{1}{4}F_{\mu\nu}F^{\mu\nu}\ +\
\frac{1}{2}i\,\bar{\psi}\gamma^\mu\partial_\mu\psi\,-\,
\frac{1}{2}i\,\partial_\mu\bar{\psi}\gamma^\mu\psi\ -\
e\bar{\psi}\gamma^\mu A_\mu\psi\ ,
\end{equation}
with $F_{\mu\nu}=\partial_\mu A_\nu-\partial_\nu A_\mu$ the gauge field 
strength, and $e$ the U(1) gauge coupling constant. In 1+1 dimensions and 
in units of mass, the gauge field $A_\mu$ is dimensionless, the spinors 
$\psi_\pm$ have dimension 1/2, and $e$ dimension unity. In covariant form, 
the equations of motion are
\begin{equation}
\gamma^\mu(i\partial_\mu-eA_\mu)\psi=0\ \ ,\ \ 
\partial^\nu\,F_{\nu\mu}=e\bar{\psi}\gamma_\mu\psi=e\,J_\mu\ .
\end{equation} 
When written out in chiral components, the Weyl spinors
$\psi_\pm(t\pm x)$ are seen to be left- and right-moving, respectively,
in the absence of any interaction, $e=0$.

On account of its antisymmetry, the field strength $F_{\mu\nu}$ represents 
a single field, $F_{01}$, which coincides with the electric field, indeed
a pseudoscalar field in 1+1 dimensions,
\begin{equation}
E=-\partial_1A^0-\partial_0A^1=F_{01}=-F^{01}\ .
\end{equation}
In particular, Gauss' law reads $\partial_1E=e\psi^\dagger\psi$.

By construction, the system is invariant under the following
global symmetries. Besides Poincar\'e invariance in the spacetime manifold,
one also has the vector U(1) and axial U(1)$_A$ symmetries whose Noether
currents are, res\-pec\-ti\-ve\-ly,
\begin{equation}
J^\mu=\bar{\psi}\gamma^\mu\psi\ \ ,\ \ 
J^\mu_5=\bar{\psi}\gamma^\mu\gamma_5\psi\ ,
\end{equation}
with the duality relations $J^\mu_5=-\epsilon^{\mu\nu}J_\nu$ and
$J^\mu=-\epsilon^{\mu\nu}J_{5\nu}$. The space integral of the time component
of each of these currents defines the corresponding Noether charges $Q$ and
$Q_5$, respectively. At the classical level, both are conserved quantities, 
thus generating a symmetry of the dynamics. For the quantized system however, 
the axial charge is no longer conserved due to quantum effects.

The U(1) vector symmetry is of course the gauge symmetry of the system,
acting as
\begin{equation}
\psi'(t,x)=e^{-ie\epsilon(t,x)}\psi(t,x)\ \ ,\ \ 
A'_\mu(t,x)=A_\mu(t,x)+\partial_\mu\epsilon(t,x)\ ,
\end{equation}
with $\epsilon(t,x)$ an arbitrary function of spacetime of the general form
\begin{equation}
\epsilon(t,x)=\epsilon_0(t,x)+\frac{2\pi k}{eL}x\ ,
\end{equation}
$\epsilon_0(t,x+L)=\epsilon_0(t,x)$ being an arbitrary periodic function
on the spatial circle, defined modulo $2\pi/|e|$ and parametrizing small U(1)
gauge transformations, while the nonperiodic contribution proportional to $x$ 
represents the homotopy class of large U(1) gauge transformations of
winding number $k$, $k$ being an arbitrary integer which labels the 
corresponding element of the gauge modular group $\Z$. These two classes of 
gauge transformations are to be included into the construction of the 
physical projector and the identification of physical states.  

\section{The Basic Hamiltonian Formulation}
\label{Sect3}

Being straightforward enough, the Hamiltonian analysis of constraints
is not detailed here. Let us only remark that the above Lagrangian density
is already in Hamiltonian form in the fermionic sector,\cite{Gov1,Gov5} 
while $A^0$ is necessarily the Lagrange multiplier for a first-class 
constraint, in fact the sole first-class constraint of the system, namely 
the generator of small U(1) gauge transformations which is nothing but 
Gauss' law.\cite{Gov1} Hence only $A^1$ needs to have its conjugate momentum 
included in order to identify the basic Hamiltonian formulation\cite{Gov1} 
of the SchM. In terms of the corresponding first-order action on phase space, 
this formulation is defined by
\begin{equation}
S=\int dt\int_0^L dx\left\{\partial_0A^1\pi_1+
\frac{1}{2}i\psi^\dagger\partial_0\psi-\frac{1}{2}i\partial_0\psi^\dagger\psi
-{\cal H}-A^0\phi+\partial_1(A^0\pi_1)\right\}\ ,
\label{eq:SH}
\end{equation}
where ${\cal H}$ is the first-class Hamiltonian density,
\begin{equation}
{\cal H}=\frac{1}{2}\pi^2_1
-\frac{1}{2}i\psi^\dagger\gamma_5(\partial_1-ieA^1)\psi
+\frac{1}{2}i(\partial_1+ieA^1)\psi^\dagger\gamma_5\psi\ ,
\label{eq:firstH}
\end{equation}
while $\phi$ is the first-class constraint of the system,
\begin{equation}
\phi=\partial_1\pi_1+e\psi^\dagger\psi\ ,
\label{eq:firstconstraint}
\end{equation}
whose Lagrange multiplier is nothing else than the time component $A^0$
of the gauge field. This first-order action encodes the 
fundamental Grassmann graded Poisson bracket structure given by
($\alpha,\beta=+,-$),
\begin{equation}
\{A^1(t,x),\pi_1(t,y)\}=\delta(x-y)\ \ ,\ \ 
\{\psi_\alpha(t,x),\psi^\dagger_\beta(t,y)\}=-i\delta_{\alpha\beta}
\delta(x-y)\ .
\end{equation}
In fact, the momentum $\pi_1$ conjugate to $A^1$ is nothing but the electric
field up to a sign, $\pi_1=F_{10}=-E$, so that the generator $\phi$ of small
Hamiltonian U(1) gauge transformations coincides with Gauss' law, one of the
two independent equations of motion in the gauge sector, namely
$\partial^\nu F_{\nu 0}=e\psi^\dagger\psi$. Manifest gauge 
invariance of the formulation readily follows from the brackets
\begin{equation}
\{\phi(t,x),{\cal H}(t,y)\}=0\ \ ,\ \ 
\{\phi(t,x),\phi(t,y)\}=0\ .
\end{equation}

Given the total Hamiltonian density 
${\cal H}_T={\cal H}+A^0\phi-\partial_1(A^0\pi_1)$, 
the Hamiltonian equations of motion read
\begin{equation}
\partial_0\psi=-\gamma_5(\partial_1-ieA^1)\psi-ieA^0\psi\ ,\
\partial_0A^1=\pi_1-\partial_1A^0\ ,\
\partial_0\pi_1=e\psi^\dagger\gamma_5\psi\ ,
\end{equation}
together with the constraint $\phi=\partial_1\pi_1+e\psi^\dagger\psi=0$.
One thus recovers the Lagrangian equation of motion for the
Dirac fermion; the equation for $A^1$ leads back to the relation
defining the conjugate momentum $\pi_1=-E$ in terms of the field $A_\mu$;
and finally the equation for $\pi_1$ reproduces the second equation of
motion in the gauge sector, namely 
$\partial^\nu F_{\nu 1}=e\psi^\dagger\gamma_5\psi$, Gauss' law being 
reproduced through the first-class constraint $\phi=0$. The dynamics of the 
system is thus equally well represented through the basic Hamiltonian 
formulation. In particular, upon substitution of the expression for $\pi_1$ 
into (\ref{eq:SH}), one recovers the original Lagrangian density, inclusive
of the explicit surface term in (\ref{eq:SH}).

Infinitesimal small U(1) gauge transformations are generated by the first-class
constraint $\phi$ through Poisson brackets. Extended to finite 
transformations, one has
\begin{equation}
\begin{array}{r c l}
\psi'(t,x)&=&e^{-ie\epsilon_0(t,x)}\psi(t,x)\ ,\\
{A^1}'(t,x)&=&A^1(t,x)-\partial_1\epsilon_0(t,x)\ ,\\
\pi'_1(t,x)&=&\pi_1(t,x)\ ,\ \\
{A^0}'(t,x)&=&A^0(t,x)+\partial_0\epsilon_0(t,x)\ ,
\end{array}
\end{equation}
$\epsilon_0(t,x)$ being an arbitrary function of periodicity $L$ in $x$
and defined modulo $2\pi/|e|$.
These transformations leave (\ref{eq:SH}) exactly invariant, and coincide 
with the small gauge transformations of the Lagrangian formulation 
of the system.

Even though large U(1) gauge transformations $\epsilon(t,x)=(2\pi kx)/(eL)$
are not generated by the first-class constraint $\phi$, nevertheless they
also define an invariance of the basic Hamiltonian action (\ref{eq:SH}),
without any surface term being induced. The phase space transformations are 
then given by
\begin{equation}
\begin{array}{r c l}
\psi'(t,x)&=&e^{-ie\epsilon(t,x)}\psi(t,x)=e^{-i\frac{2\pi k}{L}x}\psi(t,x)\ ,\\
{A^1}'(t,x)&=&A^1(t,x)-\partial_1\epsilon(t,x)=A^1(t,x)-\frac{2\pi k}{eL}\ ,\\
\pi'_1(t,x)&=&\pi_1(t,x)\ ,\ \\
{A^0}'(t,x)&=&A^0(t,x)+\partial_0\epsilon(t,x)=A^0(t,x)\ .
\end{array}
\end{equation}
From these expressions, it follows that in terms of a Fourier series
expansion of the field $A^1(t,x)$ on the circle, all its nonzero modes may 
always be gauged away entirely through an appropriate small finite gauge 
transformation which in any case leaves its zero mode invariant, 
whereas large gauge transformations affect only its zero mode which is 
thus defined modulo $(2\pi)/(|e|L)$. Hence only the $x$-independent 
mode of $A^1(t,x)$ is an actual dynamical gauge invariant physical degree of
freedom taking its values on a circle of radius 
$1/(|e|L)$.\cite{Manton,Hetrick}
As we shall see, this topological feature translates at the quantum level
in terms of $\theta$-vacua. In contradistinction, the momentum $\pi_1$
conjugate to $A^1$ is totally gauge invariant and thus physical.
Even though the $A^1$ zero mode is not exatly gauge invariant except
modulo $(2\pi)/(|e|L)$ whereas the $\pi_1$ zero mode is genuinely gauge
invariant, one may suspect that these zero modes could be conjugate to
one another. However, since the nonzero modes of $A^1$ are pure gauge, 
which are the gauge invariant physical degrees of freedom of the system that 
are conjugate to the gauge invariant nonzero modes of $\pi_1$?

\section{Nonperturbative Canonical Quantization}
\label{Sect4}

The operator quantization of the model proceeds from its basic Hamiltonian
formulation. Thus, the space of quantum states must provide a representation
space for the commutation and anticommutation relations (only the
nonvanishing (anti)commutators are given throughout)
\begin{equation}
[A^1(x),\pi_1(y)]=i\delta(x-y)\ \ ,\ \ 
\{\psi_\alpha(x),\psi^\dagger_\beta(y)\}=\delta_{\alpha\beta}\delta(x-y)\ ,
\label{eq:commutations}
\end{equation}
where it is understood that one is working within the Schr\"odinger picture
considered, say, at $t=0$, an argument which henceforth is no longer displayed.
Quantum dynamics of the system is generated through the first-class
Hamiltonian density (\ref{eq:firstH}) and constraint
(\ref{eq:firstconstraint}). However, being composite ope\-ra\-tors, these
first-class quantities require a choice of operator ordering which must
remain anomaly free and gauge invariant, namely such that the gauge 
commutation relations
\begin{equation}
[\phi(x),{\cal H}(y)]=0\ \ ,\ \ 
[\phi(x),\phi(y)]=0\ 
\end{equation}
are still obtained at the operator level. In the case of the SchM, it is
possible to meet all these requirements and construct an explicit
exact diagonalization of the quantum Hamiltonian. This is achieved
through a bosonization of the fermionic sector of degrees of 
freedom.\cite{Manton,Hetrick}

\subsection{Bosonization of the fermionic degrees of freedom}

By lack of space, we do not provide details of this construction,
but only its basic ingredients.\cite{Gaby} A quantization of
the system amounts to the specification of a quantum representation space
for the commutation and anticommutation relations (\ref{eq:commutations}).
We shall obtain the above field degrees of freedom $A^1(x)$, $\pi_1(x)$
and $\psi_\pm(x)$ as composite operators of an alternative set
of degrees of freedom whose commutation relations define the space of
quantum states. For this purpose, let us introduce the following
mode algebras,
\begin{equation}
[\varphi_n,\varphi^\dagger_m]=\delta_{n,m}\ \ ,\ \ 
[Q_\pm,p_\pm]=i\ \ ,\ \ 
[A_{\pm,n},A^\dagger_{\pm,m}]=\delta_{n,m}\ ,
\label{eq:basic}
\end{equation}
where for the $\varphi_n$ and $\varphi^\dagger_m$ modes the indices
$n$ and $m$ take all integer values, while for the operators
$A_{\pm,n}$ and $A^\dagger_{\pm,m}$ they only take strictly
positive integer values. We also have $Q^\dagger_\pm=Q_\pm$ and
$p^\dagger_\pm=p_\pm$. Furthermore, the eigenvalue spectrum of the
$Q_\pm$ operators is constrained to lie within the interval $[0,2\pi]$,
namely $Q_\pm$ stands for the coordinate of a circle of radius unity.
Consequently, the eigenvalue spectra of their conjugate operators $p_\pm$
for the Heisenberg algebras $[Q_\pm,p_\pm]=i$ are quantized as
$(n_\pm+\lambda_\pm)$, $n_\pm$ being any integers and 
$\lambda_\pm$ (mod $\Z$) being the U(1) holonomies labelling all inequivalent
representations of the Heisenberg algebra on the circle of radius 
unity.\cite{Gov6} It is clear that the complete representation space of 
the commutation
relations (\ref{eq:basic}) is the tensor product of these two Heisenberg 
algebra representations characterized by $\lambda_\pm$ together with Fock 
space representations for each of the independent modes $\varphi_n$ and 
$A_{\pm,n}$ and their adjoint operators $\varphi^\dagger_n$ and
$A^\dagger_{\pm,n}$, the former being annihilation and the latter 
creation operators.

Given the fundamental algebra (\ref{eq:basic}), let us now introduce the
following operators, for all strictly positive integer values $n\ge 1$,
\begin{equation}
\begin{array}{r c l}
q_\pm&=&Q_\pm+N_0(\varphi_0+\varphi^\dagger_0)\ ,\\
a_{+,n}&=&A_{+,n}+N_n\sqrt{n}\left(\varphi^\dagger_n+\varphi_{-n}\right)\ ,\\
a^\dagger_{+,n}&=&
A^\dagger_{+,n}+N_n\sqrt{n}\left(\varphi_n+\varphi^\dagger_{-n}\right)\ ,\\
a_{-,n}&=&A_{-,n}+N_n\sqrt{n}\left(\varphi_n+\varphi^\dagger_{-n}\right)\ ,\\
a^\dagger_{-,n}&=&
A^\dagger_{-,n}+N_n\sqrt{n}\left(\varphi^\dagger_n+\varphi_{-n}\right)\ ,\
\end{array}
\end{equation}
$N_n$ being some normalization factor to be specified presently, as well as
the bosonic fields (in the Schr\"odinger picture),
\begin{equation}
\begin{array}{r c l}
\varphi(x)&=&A\sum_{n=-\infty}^{+\infty} N_n
\left[\varphi_n\,e^{\frac{2i\pi}{L}nx}\,+\,
\varphi^\dagger_n\,e^{-\frac{2i\pi}{L}nx}\right]\ ,\\
\pi_\varphi(x)&=&\frac{-i}{2AL}\sum_{n=-\infty}^{+\infty}\frac{1}{N_n}
\left[\varphi_n\,e^{\frac{2i\pi}{L}nx}\,-\,
\varphi^\dagger_n\,e^{-\frac{2i\pi}{L}nx}\right]\ ,\\
\Phi_\pm(x)&=&Q_\pm\pm\frac{2\pi}{L}p_\pm x+
\sum_{n=1}^\infty\frac{1}{\sqrt{n}}
\left[A^\dagger_{\pm,n}\,e^{\pm\frac{2i\pi}{L}nx}\,+\,
A_{\pm,n}\,e^{\mp\frac{2i\pi}{L}nx}\right]\ ,\\
\phi_\pm(x)&=&q_\pm\pm\frac{2\pi}{L}p_\pm x+
\sum_{n=1}^\infty\frac{1}{\sqrt{n}}
\left[a^\dagger_{\pm,n}\,e^{\pm\frac{2i\pi}{L}nx}\,+\,
a_{\pm,n}\,e^{\mp\frac{2i\pi}{L}nx}\right]\ ,
\end{array}
\end{equation}
the factor $A$ being yet another normalization constant in terms of which 
$N_n$ is expressed as
\begin{equation}
N_n=\frac{1}{|A|}\,\frac{1}{\sqrt{L}\sqrt{2\omega_n}}\ ,\ \ 
\omega_n=\sqrt{\mu^2+\left(\frac{2\pi}{L}n\right)^2}\ ,
\end{equation}
where $\mu>0$ is a given mass parameter. When applied to the SchM, the 
quantities $A$ and $\mu$ are related to the gauge coupling constant $e$.
Note that we also have $\phi_\pm(x)-\Phi_\pm(x)=\varphi(x)/A$.
Only the fields $\varphi(x)$, $\pi_\varphi(x)$ and $\Phi_\pm(x)$ define 
the independent set of basic phase space degrees of freedom.

A straightforward analysis then establishes that the only nonvanishing
commutation relations for these fields are
\begin{equation}
\begin{array}{r c l}
\left[\varphi(x),\pi_\varphi(y)\right]&=&i\delta(x-y)\ ,\\
\left[\Phi_\pm(x),\pi_\varphi(y)\right]=0\ \ &,&\ \ 
\left[\Phi_\pm(x),\partial_1\Phi_\pm(y)\right]=\pm 2i\pi\delta(x-y)\ ,\\
\left[\phi_\pm(x),\pi_\varphi(y)\right]=\frac{i}{A}\delta(x-y)\ \ &,&\ \
\left[\phi_\pm(x),\partial_1\phi_\pm(y)\right]=\pm 2i\pi\delta(x-y)\ .
\end{array}
\label{eq:commutations2}
\end{equation}
In other words, $\varphi(x)$ and its conjugate momentum
$\pi_\varphi(x)$ define the degrees of freedom of a scalar field
which is periodic in $x$ on the circle of radius $R$ and takes its values
in the real line, while $\Phi_\pm(x)$ are two independent chiral bosons
each taking their values in the circle of radius unity and obeying
quantum boundary conditions such that 
$\Phi_\pm(x+L)=\Phi_\pm(x)+2\pi(n_\pm+\lambda_\pm)$, $(n_\pm+\lambda_\pm)$
being the eigenvalues of the momentum zero modes $p_\pm$, namely corresponding
to twisted chiral bosons on the circle of radius unity when 
$\lambda_\pm\ne 0$ (mod $\Z$).
These nontrivial topology properties of the chiral boson zero modes turn out 
to be crucial for the correct representations of the gauge invariance 
properties of the system under both small and large U(1) gauge transformations.

In order to make contact with the fermionic degrees of freedom, let
us now introduce the following field operators,
\begin{equation}
\begin{array}{r l}
\psi_\pm(x)=&\frac{1}{\sqrt{L}}\,e^{\pm\frac{i\pi}{L}x}\,
e^{i\frac{\pi}{2}\eta p_\mp}\,e^{\pm i\lambda q_\pm}\,
e^{\frac{2i\pi\lambda}{L}p_\pm x}\\
&\times \prod_{n=1}^\infty\,e^{\pm\frac{i\lambda}{\sqrt{n}}a^\dagger_{\pm,n}
e^{\pm\frac{2i\pi}{L}nx}}
\prod_{n=1}^\infty\,e^{\pm\frac{i\lambda}{\sqrt{n}}a_{\pm,n}
e^{\mp\frac{2i\pi}{L}nx}}\ ,
\end{array}
\label{eq:fermion}
\end{equation}
where $\eta$ and $\lambda$ are two real constants such that 
$\eta^2=1=\lambda^2$. Except for the first three factors, this
operator is nothing but the exponential $e^{\pm i\lambda\phi_\pm(x)}$
with an operator ordering such that all the $q_\pm$ and $a^\dagger_{\pm,n}$ 
are to the left of all the $p_\pm$ and $a_{\pm,n}$ operators. These fields 
possess the following holonomy properties
\begin{equation}
\psi_\pm(x+L)=-e^{2i\pi\lambda\lambda_\pm}\,\psi_\pm(x)\ ,
\end{equation}
on account of the topology properties of the $Q_\pm$ and $p_\pm$
zero modes of the chiral fields $\Phi_\pm(x)$ and $\phi_\pm(x)$.
That the fields $\psi_\pm(x)$ are indeed fermions in 1+1 dimensions follows
from their anticommutation relations obtained from the fundamental
algebra in (\ref{eq:basic}),
\begin{equation}
\{\psi_\alpha(x),\psi^\dagger_\beta(y)\}=\delta_{\alpha\beta}\delta(x-y)
\ \ ,\ \ \alpha,\beta=+,-\ .
\end{equation}
In other words, the bosonic fields $\phi_\pm(x)$
and their quantum states also provide a representation of the
fermionic algebra in (\ref{eq:commutations}), irrespective of the choice
of sign factors $\eta=\pm 1$ and $\lambda=\pm 1$. 

Hence, the fermionic matter sector of the SchM may be bosonized in 
this manner in terms of the above bosonic field degrees of freedom, provided
the U(1) holonomies $\lambda_\pm$ are chosen such that 
$\lambda_\pm=\lambda\alpha_\pm$ (mod $\Z$). Note that in order to obtain the 
anticommutation relations in (\ref{eq:commutations}) on their own,
so far it is only through the combinations 
$\phi_\pm(x)=\Phi_\pm(x)+\varphi(x)/A$ that the fields $\varphi(x)$ and 
$\Phi_\pm(x)$ are involved in this construction. However, when considered
together with the bosonic commutations relations for the gauge field
in (\ref{eq:commutations}), the whole set of degrees of freedom associated
to each of the independent fields $\varphi(x)$, $\pi_\varphi(x)$ and 
$\Phi_\pm(x)$ separately is essential to establish the complete operator 
diagonalization of the model.

\subsection{Gauge invariant point-splitting regularization}

Given the above bosonization of the fermionic fields, one must now
define the composite operators associated to the first-class Hamiltonian
density (\ref{eq:firstH}) and constraint (\ref{eq:firstconstraint}).
Since gauge invariance must be preserved at all steps, a gauge invariant
point-splitting regularization of short-distance singularities is a
relevant choice. Namely, associated for example to the product 
$\psi^\dagger_\pm(x)\psi_\pm(x)$ which is ill-defined as such as a composite
operator, one considers the gauge invariant quantity
\begin{equation}
\psi^\dagger_\pm(y)\,e^{ie\int_x^y du A^1(u)}\,\psi_\pm(x)
\end{equation}
in the limit where $y$ goes to $x$, and subtracts any divergent term
that arises. Since the Wilson line phase factor is inserted in this
point-split product of the two fields, the result is manifestly gauge invariant.
This example applies directly to the definition of the vector and axial
current operators. A similar approach is used for the products of fermionic
fields, now including the U(1) covariant derivative, that contribute to the
first-class Hamiltonian.

In terms of the bosonized representation of the fermion fields, one then 
obtains the following definitions for composite quantum operators,
\begin{equation}
\psi^\dagger_\pm\psi_\pm=-\frac{\lambda}{2\pi}
\left(\partial_1\phi_\pm\mp e\lambda A^1\right)\ ,
\end{equation}
so that the first-class constraint reads,
\begin{equation}
\phi=\partial_1\pi_1+e\psi^\dagger\psi=\partial_1\pi_1-\frac{e\lambda}{2\pi}
\partial_1(\phi_++\phi_-)\ ,
\label{eq:firstphi}
\end{equation}
while the fermion contributions to the first-class Hamiltonian density 
${\cal H}$ are
\begin{equation}
\frac{1}{2}i\psi^\dagger_\pm\left(\partial_1-ieA^1\right)\psi_\pm-
\frac{1}{2}i\left(\partial_1+ieA^1\right)\psi^\dagger_\pm\psi_\pm=
\pm\frac{1}{4\pi}\left(\partial_1\phi_\pm\mp e\lambda A^1\right)^2
\mp\frac{\pi}{12L^2}\ .
\end{equation}
Note that the vector and axial currents as well as the first-class constraint,
which are composite quantities in terms of the fermionic fields, are
local noncomposite ope\-ra\-tors in terms of the bosonic degrees of freedom, 
a property not shared, though, by the first-class Hamiltonian whose definition 
thus still requires the specification of some choice of operator ordering.

\subsection{The diagonalized operator basis}

Let us now turn to the definition of the first-class Hamiltonian
operator in correspondence with (\ref{eq:firstH}). Given the above
bosonization and gauge invariant point-splitting regularization, one finds
\begin{equation}
{\cal H}=\frac{1}{2}\pi^2_1
+\frac{1}{4\pi}\left(\partial_1\phi_+-e\lambda A^1\right)^2
+\frac{1}{4\pi}\left(\partial_1\phi_-+e\lambda A^1\right)^2\ ,
\end{equation}
where the Casimir energy contribution $-\pi/(6L^2)$ has been ignored since
the energy is defined up to an arbitrary constant anyway.
However, completing the sum of squares as follows
\begin{equation}
\begin{array}{r l}
{\cal H}=\frac{1}{2}\pi^2_1
+&\frac{1}{8\pi}\left(\partial_1\phi_++\partial_1\phi_--\frac{2\pi}{e\lambda}
\partial_1\pi_1+\frac{2\pi}{e\lambda}\partial_1\pi_1\right)^2\\
+&\frac{1}{8\pi}\left(\partial_1\phi_+-\partial_1\phi_--2e\lambda A^1\right)^2
\ ,
\end{array}
\end{equation}
one recognizes the explicit appearance of the first-class constraint
$\phi$ in the form (\ref{eq:firstphi}), so that one may also write
\begin{equation}
\begin{array}{r l}
{\cal H}=\frac{1}{2}\pi^2_1&+\frac{\pi}{2e^2}\left(\partial_1\pi_1\right)^2
+\frac{e^2}{2\pi}\left(A^1-\frac{1}{2e\lambda}\partial_1(\phi_+-\phi_-)\right)^2
\\
&+\frac{\pi}{2e^2}\phi^2-\frac{\pi}{e^2}\phi(\partial_1\pi_1)\ .
\end{array}
\end{equation}
Given this expression, let us then introduction the following definitions,
\begin{equation}
\varphi=-\frac{1}{\mu}\pi_1\ \ ,\ \ 
\pi_\varphi=\mu\left[A^1-\frac{1}{2e\lambda}\partial_1(\phi_+-\phi_-)\right]\ ,
\end{equation}
$\mu$ being the mass parameter
\begin{equation}
\mu=\frac{|e|}{\sqrt{\pi}}\ .
\end{equation}
In particular, note that except for the mass factor $\mu$, the field
$\varphi$ is nothing else than the electric field, $\varphi=E/\mu$.
The first-class Hamiltonian then reads
\begin{equation}
{\cal H}=\frac{1}{2}\pi^2_\varphi+\frac{1}{2}\left(\partial_1\varphi\right)^2
+\frac{1}{2}\mu^2\varphi^2+\frac{1}{2}\left(\frac{\phi}{\mu}\right)^2+
\left(\frac{\phi}{\mu}\right)(\partial_1\varphi)\ ,
\label{eq:H2}
\end{equation}
a most suggestive result indeed!

To confirm the fact that the fields $\varphi$ and $\pi_\varphi$ are
conjugate to one another, one need only compute their algebra given the
quantization rules (\ref{eq:commutations}), leading to the canonical 
commutation relations,
\begin{equation}
[\varphi(x),\pi_\varphi(y)]=i\delta(x-y)\ .
\end{equation}
Furthermore, the remaining commutation relations are
\begin{equation}
[\phi_\pm(x),\pi_\varphi(y)]=-\frac{i\pi\mu}{e\lambda}\delta(x-y)\ \ ,\ \ 
[\phi_\pm(x),\partial_1\phi_\pm(y)]=\pm 2i\pi\delta(x-y)\ .
\end{equation}
This set of commutations relations is thus equivalent, upon bosonization,
to the original set in (\ref{eq:commutations}). Furthermore, it is seen
to coincide exactly with the relations in (\ref{eq:commutations2}) provided
only one sets $A=-e\lambda/(\pi\mu)$ and identifies the mass scale $\mu$
with the quantity $\mu=|e|/\sqrt{\pi}$, so that 
$N_n=\sqrt{\pi}/(\sqrt{L}\sqrt{2\omega_n})$. 

Consequently, all quantities must now be expressed in terms of the mode 
degrees of freedom $Q_\pm$, $p_\pm$, $\varphi_n$, $A_{\pm,n}$ and their
adjoint operators, or equi\-va\-lent\-ly in terms of the fields $\varphi(x)$,
$\pi_\varphi(x)$ and $\Phi_\pm(x)$. Thus, the first-class constraint is 
finally given by
\begin{equation}
\phi=-\frac{e\lambda}{2\pi}\partial_1\left[\Phi_+ + \Phi_-\right]\ ,
\end{equation}
namely,
\begin{equation}
\begin{array}{r l}
\phi(x)=&-\frac{e\lambda}{L}\Big\{(p_+-p_-)+\\
& +i\sum_{n=1}^\infty\sqrt{n}\left((A^\dagger_{+,n}+A_{-,n})\,
e^{\frac{2i\pi}{L}nx}\,-\,(A^\dagger_{-,n}+A_{+,n})\,
e^{-\frac{2i\pi}{L}nx}\right)\Big\}\ ,
\end{array}
\label{eq:opphi}
\end{equation}
while the first-class Hamiltonian reads
\begin{equation}
H=\sum_{n=-\infty}^{+\infty} \omega_n\varphi^\dagger_n\varphi_n\ +\
\int_0^L dx \left[\frac{1}{2}
\left(\frac{\phi}{\mu}\right)\left(\frac{\phi}{\mu}\right)+
\left(\frac{\phi}{\mu}\right)(\partial_1\varphi)\right]\ ,
\label{eq:opH}
\end{equation}
where the contribution of the $(\varphi,\pi_\varphi)$ degrees of freedom is 
displayed ex\-pli\-ci\-tly in terms of their creation and annihilation
operators following the usual normal ordering prescription for that sector.
For the gauge degrees of freedom contribution to $H$ though, since 
the operators $\phi(x)$ and $\varphi(x)$ commute with themselves and one 
another, these operators are multiplied as shown, without any normal
ordering prescription in terms of the modes $Q_\pm$, $p_\pm$, $A_{\pm,n}$
and $A^\dagger_{\pm,n}$ being applied. The reason for this choice is
that these contributions to the total Hamiltonian then vanish identically
for all physical states, while otherwise some infinite normal ordering
constant would arise. Finally, the vector and axial charge operators 
are given by
\begin{equation}
\begin{array}{r c l}
Q&=&\int_0^L dx\,\psi^\dagger\psi=-\lambda(p_+-p_-)\ ,\\
Q_5&=&\int_0^L dx\psi^\dagger\gamma_5\psi=
-\frac{i\lambda\sqrt{2\mu}}{2\sqrt{\pi}\sqrt{L}}\,
\left[\varphi_0-\varphi^\dagger_0\right]\ .
\end{array}
\label{eq:opQ}
\end{equation}

In conclusion, the SchM and its dynamics has been quantized in terms of 
the decoupled set of basic bosonic fields, $\varphi(x)$ and $\pi_\varphi(x)$
on the one hand, and $\Phi_\pm(x)$ on the other hand, whose commutation 
relations (\ref{eq:basic}) are those of a periodic scalar field 
taking its values in the real line and two twisted chiral bosons 
taking their values in a circle of radius unity. These two sectors of 
operators, which commute with one another, see (\ref{eq:commutations2}), 
diagonalize the commutator and anticommutator algebra (\ref{eq:commutations}) 
defining the quantum system as well as its first-class Hamiltonian $H$ and 
constraint $\phi(x)$ operators, (\ref{eq:opphi}) and (\ref{eq:opH}). In 
particular, the field $\varphi(x)$ coincides, up to the mass scale $\mu$, 
with the electric field, whose nonperturbative quantum dynamics 
is that of a pseudoscalar field whose quanta have a mass $\mu$ proportional 
to the U(1) gauge coupling $e$ of the original gauge field $A_\mu$ to
the original fermionic degrees of freedom. The appearance of this screening 
length $1/\mu$ for the electric field is a purely quantum effect, and is 
one of the consequences within the physical sector of the system of its 
underlying fermionic one even though the latter degrees of freedom are
not gauge invariant except through specific combinations. The structure of 
the physical ground state is also dependent on gauge invariant topology 
features of the fermionic matter excitations. Furthermore, the phy\-si\-cal 
spectrum of U(1) gauge invariant states consists only of superpositions of the 
massive free quanta of the electric field. These results are described next.

\section{The Physical Projector and Spectrum}
\label{Sect5}

Having defined a quantization of the model, different issues may be
addressed. First, U(1) gauge invariance remains a symmetry of the quantum
dynamics, namely the first-class algebra $[\phi(x),\phi(y)]=0$ and
$[H,\phi(x)]=0$ is preserved for the quantum operators. Likewise,
the Poincar\'e algebra remains satisfied,\cite{Gaby} namely manifest spacetime
covariance is maintained through quantization on the space of quantum states,
as well as on the subspace of physical states to be constructed hereafter
since the Poincar\'e algebra commutes with the U(1) generator $\phi$.

Switching to the Heisenberg picture (as indicated by the subscript ``$H$"
in the relations hereafter), it is possible to establish the 
operator equations of motion generated by the total Hamiltonian
$H_T=H+\int_0^L dx\left[A^0\phi+\partial_1(A^0\varphi)/\mu\right]$.  
For the operators of interest, one finds
\begin{equation}
\begin{array}{r c l}
[\partial^2_0-\partial^2_1+\mu^2]\varphi_H&=&
\partial_1\left(\frac{\phi_H}{\mu}\right)\ ,\\
\partial_0\left(\frac{\phi_H}{\mu}\right)&=&0\ ,\\
\partial_\mu\,J^\mu_H=0\ \ &,&\ \ \frac{d}{dt}Q_H=0\ ,\\
\partial_\mu\,J^\mu_{5\,H}=\frac{\mu^3}{e}\varphi_H=\frac{e}{\pi}E_H\ \ &,&\ \ 
\frac{d}{dt}Q_{5 H}=\frac{e}{\pi}\int_0^L dx\, E_H\ .
\end{array}
\end{equation}
Note that the axial current $J^\mu_{5\,H}$ and charge $Q_{5\,H}$ are no
longer conserved at the quantum level as they are at the classical one,
showing that this classical global symmetry is explicitly broken by
quantum properties of the system. This chiral anomaly is directly
proportional to the electric field, namely the actual physical configuration 
space of the system. In contradistinction, the local U(1) vector symmetry is
explicitly preserved in the quantized system on account of the gauge
invariant regularization through point splitting.

Finally, let us turn to the issue of gauge invariance. It is clear that
among the basic fields, the electric field sector $(\varphi,\pi_\varphi)$
is explicitly gauge invariant, while the chiral bosons $\Phi_\pm(x)$ are not.
Given any $L$-periodic function $\epsilon_0(x)$ and an operator ${\cal O}(x)$, 
the corresponding finite small gauge transformation generated by the 
first-class constraint $\phi(x)$ is given by the adjoint action of the
following unitary operator
\begin{equation}
{\cal O}'(x)=U(\epsilon_0)\,{\cal O}(x)\,U^{-1}(\epsilon_0)\ \ ,\ \ 
U(\epsilon_0)=e^{i\int_0^L dx \epsilon_0(x)\phi(x)}\ .
\end{equation}
For the basic fields, one finds
\begin{equation}
\varphi'(x)=\varphi(x)\ \ ,\ \ 
\pi_\varphi'(x)=\pi_\varphi(x)\ \ ,\ \ 
\Phi_\pm'(x)=\Phi_\pm(x) \mp e\lambda\epsilon_0(x)\ .
\end{equation}
Since the latter relation for $\Phi_\pm(x)$ implies that the fields 
$\phi_\pm(x)$ also transform with the same rules, the correct phase 
transformation of the fermionic fields (\ref{eq:fermion}) is reproduced.
Given a mode expansion of the gauge parameter,
\begin{equation}
\epsilon_0(x)=\sum_{n=-\infty}^{+\infty}\,
\epsilon^{(0)}_n\,e^{\frac{2i\pi}{L}nx}\ ,
\end{equation}
the above rules translate into the gauge transformation for chiral modes,
\begin{equation}
Q'_\pm=Q_\pm\mp e\lambda\epsilon^{(0)}_0\ \ ,\ \ 
A'_{\pm,n}=A_{\pm,n}\mp e\lambda\sqrt{n}\,\epsilon^{(0)}_{\mp n}\ \ ,\ \ 
{A^\dagger}'_{\pm,n}=A^\dagger_{\pm,n}\pm e\lambda\sqrt{n}\,
\epsilon^{(0)}_{\pm n}\ ,
\end{equation}
showing that the chiral momentum zero modes $p_\pm$ are invariant under 
small gauge transformations, as are of course also all the modes $\varphi_n$ and
$\varphi^\dagger_n$ of the fields $\varphi(x)$ and $\pi_\varphi(x)$. 
Furthermore, recall that the gauge parameter $\epsilon_0(x)$
is defined modulo $2\pi/|e|$, hence so is its zero mode $\epsilon^{(0)}_0$.
Consequently, the actual reason for our previous assumption that the
chiral coordinate zero modes $Q_\pm$ take their values on a circle
of radius unity, namely that they are defined modulo $2\pi$, stems
directly from their properties under small U(1) gauge transformations.

Consider now a large gauge transformation of homotopy class $k$, thus
associated to the parameter $\epsilon(x)=(2\pi kx/eL)$. Since in this case
the fermion fields $\psi_\pm$ transform with the phase factor
$e^{-2i\pi kx/L}$ while the gauge field $A^1(x)$ is shifted by the quantity
$-\partial_1\epsilon(x)=-2\pi k/(eL)$, it follows that among all the modes
of the basic fields $\varphi(x)$, $\pi_\varphi(x)$ and $\Phi_\pm(x)$, only the
chiral momentum zero modes $p_\pm$ must transform under large gauge 
transformations according to the following rule
\begin{equation}
p'_\pm=p_\pm-\lambda k\ .
\end{equation}
This transformation is generated by a unitary operator of the form
\begin{equation}
U_k=C_k\,e^{ik\lambda(Q_++Q_-)}\ \ ,\ \ 
p'_\pm=U_k\,p_\pm\,U^{-1}_k\ ,
\end{equation}
where $C_k$ is a cocycle factor introduced in order to obtain the following
group composition law for any two large gauge transformations
of U(1) homotopy classes $k$ and $\ell$,
\begin{equation}
U_k\,U_\ell=U_{k+\ell}\ .
\end{equation}
The general solution to this condition is $C_k=e^{i\theta k\lambda}$
where $\theta$ is an arbitrary real constant defined modulo $2\pi$.
Hence, the operator representation of large U(1) gauge transformations of 
homotopy class $k$ is simply $U_k=e^{i\theta k\lambda}\,e^{ik\lambda(Q_++Q_-)}$,
up to any transformation $U(\epsilon_0)$.

These gauge transformations also answer a question raised pre\-vious\-ly.
The gauge invariant degrees of freedom which are conjugate to the gauge 
invariant electric field are those of the momentum $\pi_\varphi(x)$, which 
is indeed a specific combination of the original gauge dependent degrees of 
freedom which is gauge invariant under both small and large gauge
transformations. Namely, $\pi_\varphi(x)$ is essentially the difference of 
the original gauge field $A^1(x)$ with the gradient 
$\partial_1(\Phi_+-\Phi_-)(x)$ of the difference of the chiral bosons.
In particular, because of this gradient contribution, the zero mode of the
electric field, which is proportional to $(\varphi_0+\varphi^\dagger_0)$,
is conjugate to the combination $(A^1_0-\pi(p_++p_-)/(e\lambda L))$, 
which is proportional to $(\varphi_0-\varphi^\dagger_0)$, rather than to
the $A^1(x)$ zero mode $A^1_0$ on its own as might have been anticipated. 
This situation is in contradistinction with results obtained within a gauge 
fixed formulation.\cite{Manton,Hetrick} Moreover, the fact that this zero mode 
$A^1_0$ is defined modulo $(2\pi)/(|e|L)$ because of its properties
under large gauge transformations, is now seen to correspond to the
analogous property for the chiral momentum zero modes $p_\pm$ which
are defined modulo integers for the same reason, since the combination
$(\varphi_0-\varphi^\dagger_0)$ is invariant under both small and large
gauge transformations. Note that the actual phase space topology of the 
twisted chiral zero mode sector $(Q_\pm,p_\pm)$ is that of a two-torus for 
each chiral boson.

Having identified the operators $U(\epsilon_0)$ and $U_k$ which generate
all small and large gauge transformations on the space of quantum states,
the physical projector is readily constructed as the sum of all these
transformations,\cite{Klauder} in order to retain only the totally gauge 
invariant components of any state,\cite{Gov1} namely the actual physical 
states of the system. Hence, the physical projector is given by
\begin{equation}
\proj=\sum_{k=-\infty}^{+\infty}\,\int_0^{\frac{2\pi}{|e|}}d\epsilon^{(0)}_0
\prod_{n=1}^\infty\int_{-\infty}^\infty d({\rm Re}\,\epsilon^{(0)}_n)
\int_{-\infty}^\infty d({\rm Im}\ \epsilon^{(0)}_n)\,
U_k\,U(\epsilon_0)\ ,
\end{equation}
where the absolute normalizations of these integrals are not specified
at this stage, the operator $\proj$ being a non-normalizable projector density.

In order to describe the physical spectrum, let us now introduce a basis
for the representation space of the algebra (\ref{eq:basic}). In the
electric field sector, we have the normalized Fock vacuum $|0>_\varphi$ 
which is annihilated by the modes $\varphi_n$ and
which is acted on with the creation operators $\varphi^\dagger_n$ to generate 
the basis of normalized states,
\begin{equation}
|\{k_p\}>_\varphi=\prod_{p=-\infty}^{+\infty}
\frac{1}{\sqrt{k_p!}}\left(\varphi^\dagger_p\right)^{k_p}\,
|0>_\varphi\ ,
\end{equation}
in which $k_p\ge 0$ quanta of integer momentum $-\infty<p<+\infty$ are 
excited. In the chiral boson sector, the creation operators 
$A^\dagger_{\pm,n}$ act on a normalized Fock vacuum $|0>_\Phi$ which is 
annihilated by the operators $A_{\pm,n}$, to also generate the full Fock 
basis of quantum excitations of the chiral fields. Finally, the chiral 
zero mode sector $(Q_\pm,p_\pm)$ is represented in the orthonormalized 
momentum eigenbasis $|n_+,n_->$, 
$<n'_+,n'_-|n_+,n_->=\delta_{n'_+,n_+}\delta_{n'_-,n_-}$, such that,
\begin{equation}
p_\pm|n_+,n_->=(n_\pm+\lambda_\pm)|n_+,n_->\ .
\end{equation}

From the explicit expression for the physical projector, it is then possible
to work out its action on this basis of the quantum space of states.\cite{Gaby}
One then finds
\begin{equation}
\proj=\one_\varphi\otimes|\Omega_\theta><\Omega_\theta|\ ,
\end{equation}
where the tensor product refers to that between the
electric and chiral sectors of degrees of freedom, while the gauge
invariant vacuum $|\Omega_\theta>$ in the sector of gauge dependent degrees
of freedom is given by the non-normalizable state
\begin{equation}
|\Omega_\theta>=\sum_{\ell=-\infty}^{+\infty}\,e^{i\ell\theta}\,
e^{-\sum_{n=1}^\infty A^\dagger_{+,n}A^\dagger_{-,n}}\,
|n_+=\ell,n_-=\ell;0>_\Phi\ ,
\label{eq:Thetavac}
\end{equation}
which is thus seen to correspond to a $\theta$-vacuum associated to
a coherent superposition, characterized by the parameter
$\theta$, of the chiral momentum eigenstates. In fact, gauge invariance
enforces the constraint $p_+=p_-$, which not only requires identical
eigenvalues $n_+=\ell=n_-$ but also identical values modulo integers for 
the U(1) holonomies $\lambda_+=\lambda_-$ (mod $\Z$), namely the parity 
invariant condition $\alpha_+=\alpha=\alpha_-$ (mod $\Z$) since 
$\alpha_\pm=\lambda\lambda_\pm$ (mod $\Z$).

Consequently, all physical states of the system are spanned by the
free massive quanta of the electric field, namely the multiparticle states
\begin{equation}
|\{k_p\}>_\varphi\otimes|\Omega_\theta>\ ,
\end{equation}
which are all invariant under the action of both small and large U(1) gauge
transformations, $U(\epsilon_0)$ and $U_k$. In particular, the first-class 
constraint ope\-ra\-tor $\phi(x)$ annihilates any physical state, since
$\phi(x)|\Omega_\theta>=0$.
This basis of states also diagonalizes the total Hamiltonian $H_T$ of the
system, which thus coincides with the first-class one $H$ on the subspace
of physical states, leading to the physical energy spectrum
\begin{equation}
E(\{k_p\})=\sum_{p=-\infty}^{+\infty}\,k_p\,\omega_p\ \ ,\ \ 
\omega_p=\sqrt{\left(\frac{2\pi}{L}p\right)^2+\mu^2}\ \ ,\ \ 
\mu=\frac{|e|}{\sqrt{\pi}}\ ,
\end{equation}
whose spacetime momentum value is also
\begin{equation}
P^1(\{k_p\})=\sum_{p=-\infty}^{+\infty}\,k_p\,\left(\frac{2\pi p}{L}\right)\ .
\end{equation}
Finally, note that the non-normalizable character of physical states resides
only within the gauge invariant contribution of the gauge dependent degrees
of freedom, namely within the factorized single gauge invariant
$\theta$-vacuum $|\Omega_\theta>$.

For the sake of illustration, let us also consider the expectation values 
of some quantities of interest. Given any multiparticle state, the
expectation value of the electric field is readily seen to vanish
identically at all times, with nonvanishing fluctuations around that 
mean value however,
\begin{equation}
<E(x)>=0\ \ \ ,\ \ \
<:E^2(x):>=\frac{1}{L}\sum_{p=-\infty}^{+\infty}\,
\frac{\mu\,k_p}{\sqrt{(\frac{2\pi}{L}p)^2+\mu^2}}\ .
\end{equation}
On the other hand, the expectation value of the gauge potential $A^1(x)$
within the same multiparticle states reads,
\begin{equation}
<A^1(x)>=\frac{2\pi}{e\lambda L}\lambda_\pm=\frac{2\pi}{eL}\alpha_\pm\ .
\end{equation}
Even though the operator $A^1(x)$ is not gauge invariant, its expectation
value for gauge invariant states should be; nonetheless, it need not 
necessarily vanish for that reason. Indeed, first note that on account of 
translation invariance in $x$, the quantity $<A^1(x)>$ measures only the 
expectation value of the zero mode of $A^1(x)$, which is thus certainly 
invariant under small U(1) gauge transformations. Furthermore, because of large
U(1) gauge transformations, this zero mode is actually defined only modulo
$(2\pi)/(|e|L)$. This is indeed also the property shared by the above
result for $<A^1(x)>$, given that the U(1) holonomies $\lambda_\pm$, 
which are directly related to the choice of boundary conditions of the 
fermionic fields $\psi_\pm(x)$ in $x$, are defined modulo integers. 
Consequently, even though nonvanishing, the expectation value $<A^1(x)>$
is perfectly gauge invariant, a possibility which arises since the
zero mode of that operator is defined on a circle of radius $1/(|e|L)$
rather than the real line. Note that consistency of this result requires once 
again the two U(1) holonomies $\lambda_\pm$, or the two factors $\alpha_\pm$,
to be equal to one another modulo integers. It is interesting to remark that 
the expectation value $<A^1(x)>$ is also a measure of the fermionic boundary 
conditions. In contradistinction, the expectation value of the other original 
gauge dependent fields, namely the Weyl spinors $\psi_\pm(x)$, always vanishes
for multiparticle states, $<\psi_\pm(x)>=0$, which is again a result
perfectly consistent with gauge invariance. Since these degrees of 
freedom vary continuously for small as well as large U(1) gauge 
transformations in contradistinction to the zero mode of $A^1(x)$, being
gauge invariant necessarily their expectation value for gauge invariant 
states much vanish identically.

Note that the U(1) charge operator $Q$, see (\ref{eq:opQ}), 
vanishes identically for any physical state, on account of gauge invariance
under small gauge transformations of these states,
$Q|\{k_p\}>_\varphi\otimes|\Omega_\theta>=0$. None of the physical states
thus carries any electric charge. This conclusion includes the
physical ground state $|0>_\varphi\otimes|\Omega_\theta>$, showing
that the exact U(1) vector gauge symmetry of the quantized system
is not spontaneously broken. Had the global axial symmetry U(1)$_A$
not be broken explicitly by a quantum anomaly, the issue of the Wigner or 
Goldstone mode realization of that symmetry could have been raised as well. 
However, the quantum axial charge $Q_5$, see (\ref{eq:opQ}), does not commute
with the quantum Hamiltonian $H$, so that its action does not conserve
energy even though it maps physical states into physical states since
it commutes with the first-class constraint operator $\phi(x)$. For
instance, acting on the physical ground state, one finds
\begin{equation}
Q_5|0>_\varphi\otimes|\Omega_\theta>=
i\frac{\lambda\sqrt{2\mu}}{2\sqrt{\pi}\sqrt{L}}\ 
|k_0=1>_\varphi\otimes|\Omega_\theta>\ ,
\end{equation}
resulting in a zero momentum 1-particle excitation of the electric field. 
For the same reason, the expectation value of the chiral charge vanishes 
identically for any multiparticle state, including the physical 
ground state, $<Q_5>=0$. Furthermore, any finite axial transformation acts as 
follows on the physical ground state,
\begin{equation}
e^{i\alpha Q_5}\,|0>_\varphi\otimes|\Omega_\theta>=
|z_0>_\varphi\otimes|\Omega_\theta>\ \ \ \ \ 
{\rm with}\ \ \ \ \ z_0=-\frac{\alpha\lambda\sqrt{2\mu}}{2\sqrt{\pi}\sqrt{L}}\ ,
\end{equation}
where $|z_0>_\varphi$ stands for the holomorphic zero momentum
coherent state in the $(\varphi(x),\pi_\varphi(x))$ sector defined by
\begin{equation}
|z_0>_\varphi=e^{-\frac{1}{2}|z_0|^2}\,e^{z_0\varphi^\dagger_0}\,|0>_\varphi\ ,
\end{equation}
$z_0$ being an arbitrary complex variable.
Consequently, axial transformations act by producing coherent
zero momentum excitations of the electric field. However, they do not induce
a change in the $\theta$ parameter of the $\theta$-vacuum $|\Omega_\theta>$
in the sector of gauge dependent degrees of freedom.
This conclusion, as well as some of the above results, should be
contrasted with those available in the literature following any
gauge fixing procedure,\cite{SchM1,Manton,Hetrick} some of which are thus 
seen to be actually artefacts due to gauge fixing.

\section{Concluding Remarks}
\label{Sect6}

Through quantization but without any gauge fixing procedure
whatsoever, the original gauge and fermionic degrees of freedom of the massless
Schwinger model have been reorganized into a diagonalized set of basic gauge 
in\-va\-riant and gauge dependent operators. The gauge invariant ones are
those of the electric field whose conjugate momentum variable is the single 
gauge in\-va\-riant combination of the vector field $A^1$ with the bosonized 
Weyl spinors. The gauge dependent ones are those of the specific combinations 
of the bosonized Weyl spinors with the electric field which are decoupled from 
the gauge invariant sector. The gauge invariant dynamics and physical 
content of the system is that of a free pseudoscalar field of mass 
$\mu=|e|/\sqrt{\pi}$, namely the electric field, which may be excited in 
any multiparticle state. The gauge dependent degrees of freedom contribute
only through a uniquely defined gauge in\-va\-riant combination which is the
$\theta$-vacuum $|\Omega_\theta>$. This state is comprised, 
see (\ref{eq:Thetavac}), on the one hand,
of a condensate which is the coherent superposition of pairs of nonzero mode 
excitations of the chiral bosons $\Phi_+(x)$ and $\Phi_-(x)$ of vanishing 
total momentum, and on the other hand, of a $\theta$-vacuum coherent
superposition of the chiral boson zero mode momentum eigenstates sharing
a common eigenvalue. 

The latter feature stems from the gauge invariant
topological properties of the zero modes of the gauge dependent sector
under both small and large U(1) gauge transformations. The $\theta$ variable,
defined modulo $2\pi$, parametrizes the sole freedom left at the quantum
level in the gauge invariant physical contribution of the gauge dependent
sector. Up to that parameter, which labels all possible distinct quantum
representations of the modular group $\Z$ of large U(1) gauge transformations, 
there exists only a single gauge in\-va\-riant quantum state within the sector 
of gauge dependent degrees of freedom, namely the $\theta$-vacuum 
$|\Omega_\theta>$. All the gauge invariant topological features of the system 
thus reside in that $\theta$-vacuum for the gauge invariant combination of 
its zero modes.

The quantization and solution of the model is nonperturbative in that taking 
the limit of a vanishing gauge coupling constant $e$ leads to a singular 
behaviour. There is no analytic deformation possible of the set of field 
operators that diagonalize the interaction-free system into that which 
dia\-go\-na\-li\-zes the interacting system, thus precluding any perturbation 
expansion. Note also that the compactification of the spatial coordinate into 
a circle is essential in bringing to the fore all the topology features
of the system within its zero mode sector which are implied by gauge
invariance under both small and large U(1) gauge transformations. The radius 
$R$ or circumference $L$ of that circle provides a regularization of the 
system which allows for a discrete spectrum of quantum states free of infrared
divergences in 1+1 dimensions, and thus in particular a clear separation
of the zero mode contributions and their topological properties.

The physical projector, free of the necessity of any gauge fixing,
is thus able to unambiguously identify, through a nonperturbative quantization
which diagonalizes the dynamics of the massless Schwinger model, and 
in agreement with results established following whatever gauge fixing 
procedure at least in as far as the physical spectrum is concerned, the actual 
physical content of the model, without suffering, however, from artefacts 
resulting from gauge fixing which render the physical interpretation 
sometimes confusing. The physical projector has also achieved the 
quantization of a compact phase space, namely the product of the two two-tori 
corresponding to the zero mode sectors of the two gauge dependent chiral 
bosons, without having recourse to any geometric quantization methods beyond 
the usual rules of canonical quantization.

\section*{Acknowledgements}

G.Y.H.A. acknowledges the financial support of the ``Coop\'eration
Universitaire au D\'evelop\-pe\-ment, Commission Interuniversitaire
Francophone" (CUD-CIUF) of the Belgian French Speaking Community, and wishes
to thank the Institute of Nuclear Physics (Catholic University
of Louvain, Belgium) for its hos\-pi\-ta\-li\-ty while this work was being
pursued.

\end{document}